\documentclass[twocolumn, secnumarabic, nobibnotes, aps, prd]{revtex4-2}

\usepackage{amsmath}
\usepackage{amsfonts}
\usepackage{amssymb}
\usepackage{graphicx}
\usepackage{soul}

\usepackage[utf8]{inputenc}
\begin{document}

\title{Bowstring effect as a trigger for flux instabilities in thin superconductors}
\author{Leonid Burlachkov}
\email{leonid@biu.ac.il}
\author{Nikita Fuzailov}
\email{fuzailovnikita@gmail.com}
\affiliation{Department of Physics, Bar-Ilan University, Ramat-Gan 5290002, Israel}

\begin{abstract}
Magnetic vortices resemble bowstrings stretched across a corner at the initial stage of their penetration into a flat superconducting sample of a rectangular cross-section. As the external magnetic field $H_e$ reaches the threshold level $H_{th}$, a bowstring is "released" and instantaneously contracted in length with a substantial heat generation. This heat can serve as a trigger for nucleation of a flux instability (avalanche).  At $H_e \approx \sqrt{2}H_{th}$ a usual vortex penetration starts at flat edges, and the bowstring mechanism is no longer effective. We describe the geometry of bowstring-like vortices, find $H_{th}$ for disk and strip shaped superconductors as a function of their thickness to width ratio, and determine the heat effect related to a bowstring release. Our results enable a novel treatment of numerous experimental data on flux instabilities and avalanche type penetration in flat superconducting samples. A moderate anisotropy of a superconducting penetration depth ($\lambda_c > \lambda_{ab}$) diminishes or even completely removes the bowstring effect. This explains the absence of spontaneous instabilities in epitaxial $\mathrm{YBa}_{2}\mathrm{Cu}_{3}\mathrm{O}_{7-\delta }$ films at all temperatures and in $\mathrm{MgB_{2}}$ above 10\,K. 
\end{abstract}

\maketitle

A avalanche-like penetration of magnetic flux into thin superconducting samples was first reported in 1967 \cite{Wertheimer} and has attracted revived attention last three decades. As an external magnetic field $H_e$, directed perpendicular to the sample surface, exceeds a threshold level $H_{th}$, flux lightnings, or dendrites, or "branching trees" start to spontaneously propagate inside from the edges. Magneto-optical technique enables visualization of these fascinating structures in $\mathrm{Nb}$ \cite%
{Duran,Altshuler1,Alvarez,Brisbois,Jiang,Vlasko-Vlasov}, $\mathrm{NbN}$ \cite{Rudnev1,Mikheenko1,Baruch-El1,Qureishy}, $\mathrm{Nb}_{3}\mathrm{Sn}$ \cite{Rudnev2}, $\mathrm{MgB}_{2}$ \cite{Johansen,Barkov,Shantsev,Denisov1,Olsen1,Albrecht,Jing1,Jing2} and other relatively "low-$\mathrm{T}_{c}$" superconductors, see Refs.\,\cite{Altshuler2,Vestgarden,Colauto} as reviews. Such instabilities appear as well in $\mathrm{Nb}$  and $\mathrm{MgB_{2}}$ flat rings and result in a staircase dependence of a trapped flux on $H_e$ \cite{Olsen2,Shvartzberg,Jiang2,Burlachkov1}. In what concerns "high-$\mathrm{T}_{c}$" compounds, a non-equilibrium penetration was observed in $\mathrm{YBa}_{2}\mathrm{Cu}_{3}\mathrm{O}_{7-\delta }$ (YBCO) films \cite{Bujok,Leiderer,Bolz,Baziljevich,Baruch-El2,Qviller,Zhou}. However, unlike "low-$\mathrm{T}_{c}$" materials, YBCO samples require special "ignition" for appearance of flux instabilities. Such a triggering can be achieved by laser spotting \cite{Bujok,Leiderer,Bolz,Zhou} or by applying fast ramping rates $dH_{e}/dt$ \cite{Baziljevich,Baruch-El2,Qviller}.

Flux instabilities in superconductors have a lot in common with formation of discharge channels at dielectric breakdown \cite{Niemeyer}, and can be considered in terms of the theory of self-organized criticality (SOC) \cite{Bak,Wijngaarden}. However, such a description needs a Bean critical state \cite{Bean} ("sand-pile" in terms of SOC) to be shaped before an avalanche occurs, whereas magnetic lightning or dendrites usually propagate into a "empty" sample. Therefore, the most consistent description of superconducting avalanches is based on a thermomagnetic effect \cite{Aranson1,Aranson2,Denisov2}, see Ref.\,\cite{Vestgarden} as a review. Similar dendritic growth in non-equilibrium thermal conditions is commonly known in metallurgy as molten metal solidifies. If heat produced by moving vortices has no opportunity to dissipate fast enough, temperature in the channel where vortices are moving increases, flux velocity grows which leads to an enhanced heat generation. Such a positive feedback results in lightning or dendritic non-equilibrium flux propagation.

Coupled equations of flux diffusion and heat dissipation perfectly describe flux dynamics in avalanche-like structures \cite{Aranson1,Aranson2,Denisov2}. However, experimental studies testify that appearance of such an instability is an activation process, and at the stage of "nucleation" an avalanche has to overcome a surface barrier. When the barrier becomes surmountable at a distinct point, a thermodynamic pressure of the external field pushes vortices inside with an energy micro-burst, which serves as an ignition for further instability evolution. This problem of avalanche triggering at the surface has not been addressed in proper detail so far.

There are two surface hurdles for flux penetration into type-II superconductors: a Bean-Livingston barrier \cite{BL} and a geometrical one \cite{geom}. But neither of them is suitable to explain the origin of the ignition energy for an instability nucleation. Bean-Livingston barrier proves to be effective in superconductors with high values of the Ginzburg-Landau parameter $\kappa =\lambda /\xi $, where $\lambda $ and $\xi $ are the magnetic penetration depth and the coherence length, respectively. Thus, the best candidates for observation the Bean-Livingston barrier are high-$\mathrm{T}_{c}$ compounds like YBCO, whereas in $\mathrm{Nb}$ with $\kappa \simeq 1$ this effect should be absent. The experimental situation is just the opposite: flux avalanches are most common in low $\kappa $ superconductors and require special triggering in YBCO. Moreover, surmounting the Bean-Livingston barrier by a vortex \cite{BurlachkovBL} is not accompanied by any burst-like energy release. 

In a flat sample of a rectangular cross-section (stripe or disk) the field near the edges $H_{edge}\simeq \sqrt{a/s}H_{e}$ in the Meissner state, where $s$ is the sample thickness and $a$ is radius of a disk or half width of a strip. Geometric barrier \cite{geom} means that at $H_{edge}=H_{c1}$, where $H_{c1}$ is the first critical field, vortices start to penetrate and are pushed toward the sample center by the surface Meissner currents. If pinning is weak enough, penetrated vortices concentrate around the center leaving the sample periphery free of flux. Of course, such a flux motion is accompanied with a heat release. However, penetration should start at $H_{edge}=H_{c1}$ across the whole sample perimeter and in the form of a wide flux front, whereas instabilities first appear at a few distinct points. Thus geometrical barrier is not a likely candidate to describe an instability nucleation and provide a sufficient "ignition" energy for it. 

In this paper we analyze the initial stage of vortex penetration into a thin superconducting sample of a rectangular cross-section and prove the existence of a "bowstring" effect, which is effective at $H_{edge}<H_{c1}$. Consider a superconducting disk of radius $a$ or stripe of half-width $a$ embedded into a perpendicular magnetic field $H_{e}$ as shown in Fig.\,1. For a sample of zero thickness in a Meissner state, the distribution of a shielding current $j(r)$ (at $r<a$) and magnetic field $H(r)$ in its plane (at $r>a$) is \cite{Larkin,Mikheenko2}%
\begin{equation}
\begin{split}
  j(r)=\frac{H_{e}c}{\pi ^{2}}\frac{r}{\sqrt{a^{2}-r^{2}}},
  \\
  H(r)=\frac{2H_{e}}{\pi }\left( \frac{a}{\sqrt{r^{2}-a^{2}}}+\frac{\pi }{2}-\arcsin \frac{a}{r}\right)
\end{split}
\label{eq-disk-ex}
\end{equation}%
for a disk, and
\begin{equation}
j(r)=\frac{H_{e}c}{2\pi }\frac{r}{\sqrt{a^{2}-r^{2}}};\quad H(r)=\frac{H_{e}a}{\sqrt{r^{2}-a^{2}}}
\label{eq-strip-ex}
\end{equation}%
for a strip, where $c$ is the speed of light. Eqs.\,(\ref{eq-disk-ex}) and (\ref{eq-strip-ex}) are mathematically exact at $s=0$, where $s$ is the sample thickness. In a real sample of finite $s$, where $a\gg s\gg \lambda $, the surface current in the vicinity of its edge is 
\begin{equation}
j_{surf}(x)=\frac{H_{e}c}{4\pi }q\sqrt{\frac{a}{x}}, \label{eq-jsurf}
\end{equation}%
where $x=a-r\ll a$, see Fig.\,1, $q=1/\sqrt{2}$ for a strip and $q=\sqrt{2}/\pi$ for a disk, compare to Eqs.\,(\ref{eq-disk-ex}) and (\ref{eq-strip-ex}). 
\begin{figure}[t!]
    \centering
    \includegraphics[width=0.48\textwidth]{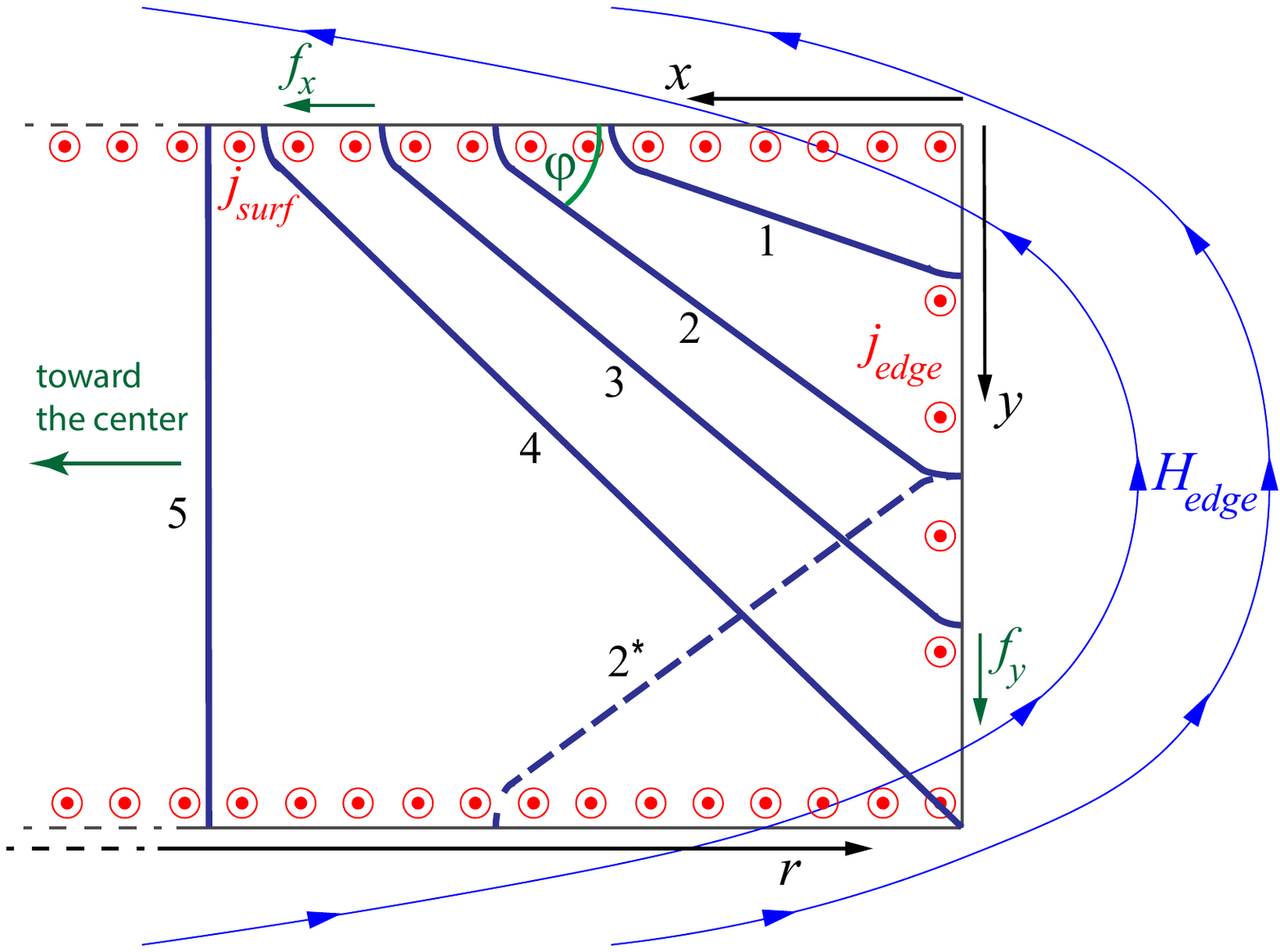}
    \caption{Bowstring vortices stretched across the corner by the Meissner currents $j_{surf}$ and $j_{edge}$. A "half" bowstring (line 2) can merge with a symmetric "half" (line 2*) and get contracted directly to a "released" state (line 5), but probability of such merging can be low. A "full" bowstring (line 4) is reached at $H_{th}$ and is instantaneously released with a heat effect.}
\end{figure}
Note that $j_{surf}$ is a linear (integrated over the $\lambda $ surface layer) current density measured in $\mathrm{A/m}$, and, of course, $j_{surf}=j/2$, since each (upper and lower) sample surface carries one half of the total Meissner current. Mathematical divergence of $H(r)$ at $r \rightarrow a$ should be cut-off at $a-r\simeq s$, thus we get from Eqs.\,(\ref{eq-disk-ex}) and (\ref{eq-strip-ex}):
\begin{equation}
H_{edge}\approx \frac{H_{e}}{\pi }q\sqrt{\frac{a}{s}}\mathrm{.}
\label{eq-Hedge}
\end{equation}%
Correspondingly, the edge Meissner current (see Fig.\,1) is%
\begin{equation}
j_{edge}\approx \frac{cH_{e}}{4\pi }q\sqrt{\frac{a}{s}}.  \label{eq-jedge}
\end{equation}

At $H_{edge}=H_{c1}$, which, as follows from Eq.\,(\ref{eq-Hedge}), is reached at%
\begin{equation}
H_{e}=H_{max} \approx \frac{\pi H_{c1}}{q}\sqrt{\frac{s}{a}}\mathrm{,}
\label{eq-Hmax}
\end{equation}%
vortices start to penetrate from the edge and then are driven toward the sample center by $j_{surf}$ \cite{geom}. However, partial penetration at the corners, where $H_{edge}$ reaches maximum, starts at $H_{e}<H_{max}$. Vortices are stretched like bowstrings across a corner by the surface and edge shielding currents, see Fig.\,1. As $H_e$ grows up, bowstrings become longer and finally reach the opposite corner at $H_{e}=H_{th}$, see line 4 in Fig.\,1. When it happens, a bowstring is released, gets instantaneously contracted in length (see line 5 in Fig.\,1) and looses energy which is transformed to heat. Let us find $H_{th}$.

A vortex, inclined arbitrary in the bulk, always "touches" a sample surface at a right angle ($90^{0}$) as shown in Fig.\,1, since currents should be parallel to a surface in the layer of depth $\lambda$. A shielding current $j$ applies the Lorentz force $f=\phi _{0}j/c$ at a vortex "$\lambda $-tip", where $\phi _{0}=\pi \hbar c/e$ is the unit flux. The first critical field $H_{c1}=4\pi \varepsilon /\phi _{0}$, where $\varepsilon \simeq (\phi _{0}/4\pi \lambda )^{2}$ is a vortex line tension, is usually determined by the condition $\varepsilon -\phi _{0}H_{e}/4\pi <0$, which means that the Gibbs free energy is reduced due to vortex formation. Alternatively, one gets the same definition for $H_{c1}$ considering the Lorentz force which stretches a vortex "nucleus" near a surface as shown in Fig.\,2. 
\begin{figure}[t!]
    \centering
    \includegraphics[width=0.48\textwidth]{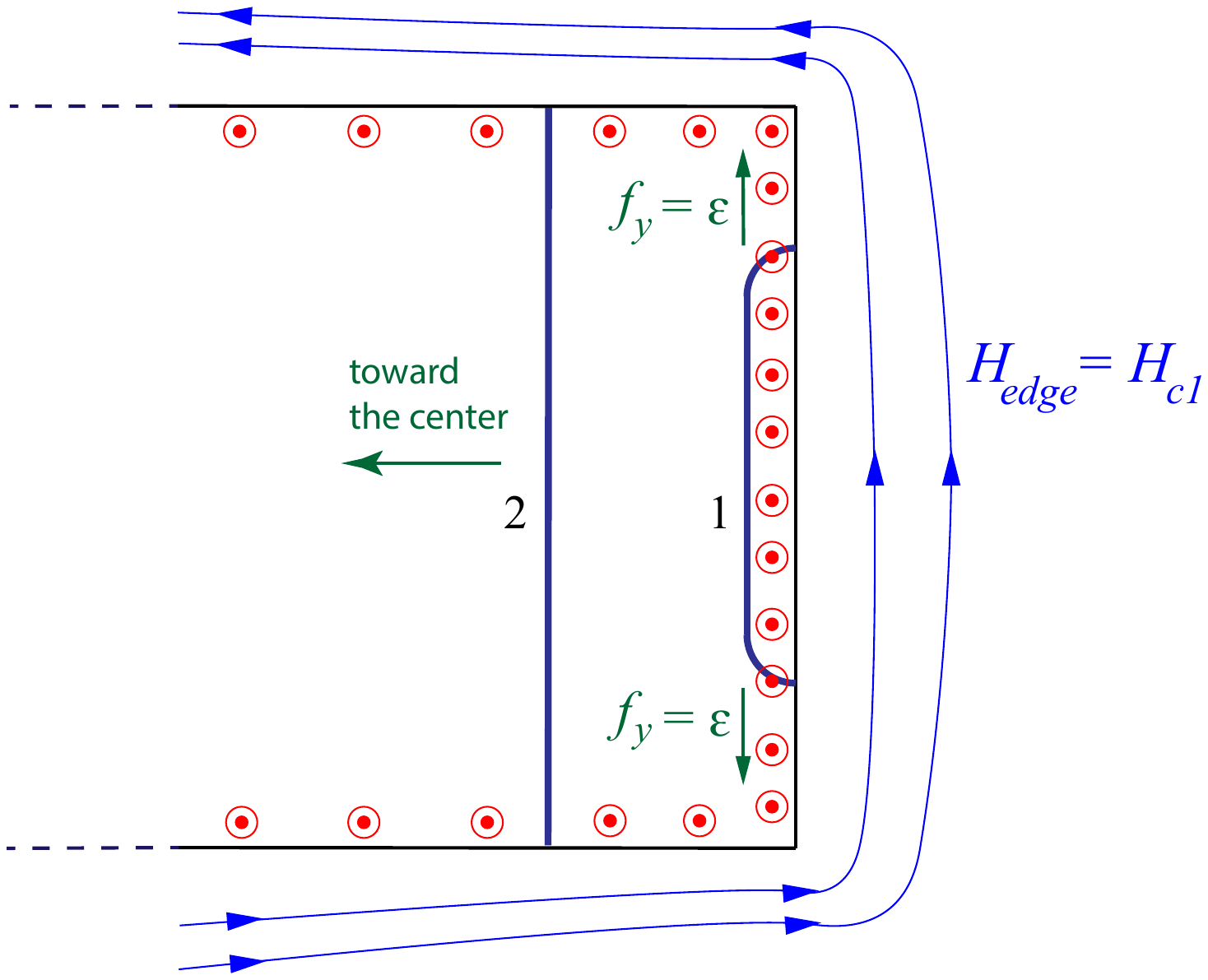}
    \caption{Usual flux penetration ("geometric barrier") at $H_e=H_{max}$. Vortices are nucleated at the flat edge (line 1) and then propagate inside (line 2) in a form of a wide flux front. The bowstring mechanism is no longer effective.}
\end{figure}
Taking into account that an edge shielding current is $j \approx cH_{e}/4\pi $ (in a cylindrical sample this equation is exact), the condition $f>\varepsilon $ returns that same value for $H_{c1}$ as obtained by its usual definition. In the same way we calculate $H_{th}$ below. 

A bowstring stays put in equilibrium provided the vortex tension is compensated by the Lorentz forces at both surface and edge: 
\begin{equation}
f_x=\frac{\phi _{0}j_{surf}}{c}=\varepsilon \cos \varphi ,\quad
f_y=\frac{\phi _{0}j_{edge}}{c}=\varepsilon \sin \varphi,  \label{eq-f}
\end{equation}%
where $\varphi $ is the angle between a bowstring "body" (excluding its $\lambda $-tips) and a surface, see Fig.\,1. If $x$ and $y$ are the coordinates of the bowstring tips as shown in Fig.\,1, the critical configuration, where $y=x\tan \varphi =s$, see line 4 in Fig.\,1, is determined by a system of three equations:%
\begin{equation}
h\sqrt{s/x}=\cos \varphi ;\quad h=\sin \varphi ;\quad x\tan \varphi =s,  \label{eq-system}
\end{equation}%
where $h=H_{e}/H_{max}=H_{edge}/H_{c1}$. An obvious solution of Eq.\,(\ref{eq-system}) is 
\begin{equation}
x=s;\quad \varphi =45^{\circ };\quad h_{th}=1/\sqrt{2}.
\label{eq-full}
\end{equation}
A usual flux penetration by nucleation of vortices at flat edges starts at $H_{edge}=H_{c1}$ (i.e., at $h=1$, $H_{e}=H_{max}$). But, as follows from Eq.\,(\ref{eq-full}), a burst-like release of bowstrings from a corner begins earlier, at $h=h_{th}$, i.e., at $H_{e}=H_{th}=H_{max}/\sqrt{2}$ for both disk or strip. Undoubtedly, this is the case for other flat sample shapes such as squares. In Fig.\,3 
\begin{figure}[t!]
    \centering
    \includegraphics[width=0.48\textwidth]{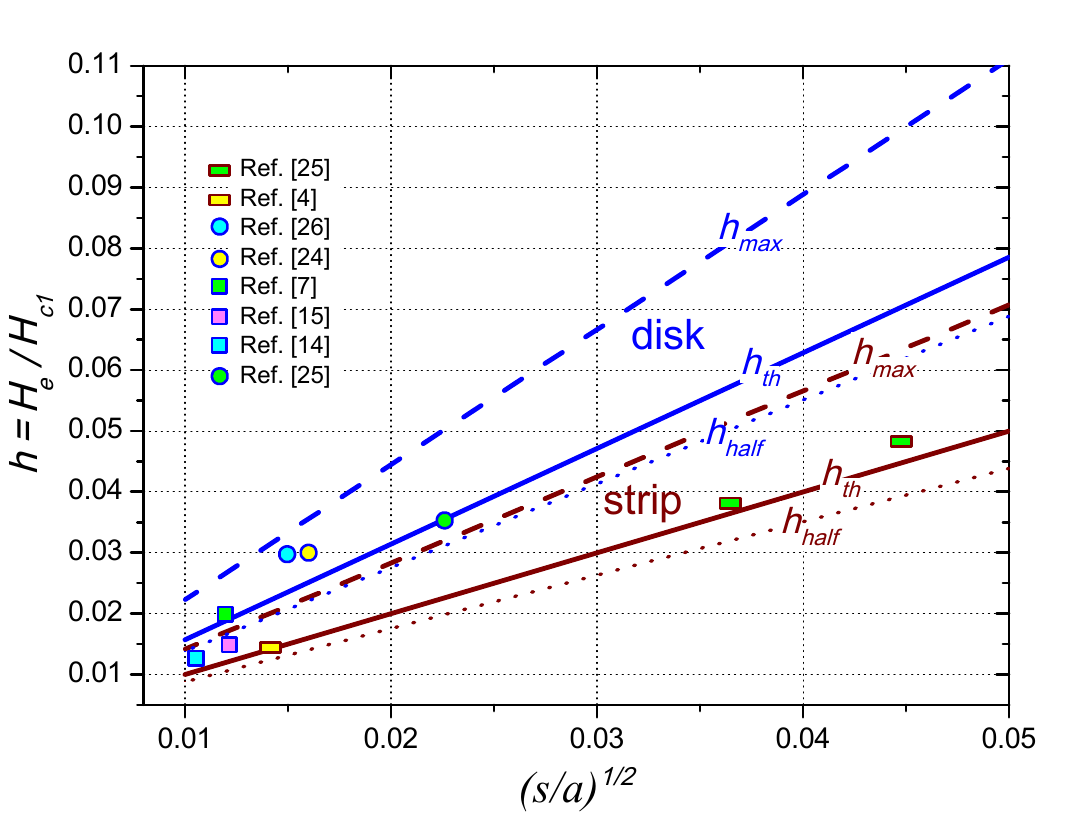}
    \caption{The fields $H_{th}$, $H_{max}$ and $H_{half}$, normalized by $H_{c1}$ for disks and strips. Avalanches should appear between $H_{th}$ (or maybe $H_{half}$) and $H_{max}$. A usual flux front penetration is expected above $H_{max}$. Available experimental results for $H_{th}$ are shown as rectangles (strips), circles (disks) and squares (square samples). The latter are treated as disks of the same area. }
\end{figure}
we plot $h_{th}$ and $h_{max}$ as functions of $\sqrt{(s/a)}$ together with experimentally found results for $h_{th}$. If a square sample with side $b$ was used in the experiment, we compare $h_{th}$ with that for a disk of the same area: $b^2=\pi a^2$. Each bowstring release from a fully stretched state (from line 4 to line 5 in Fig.\,1) results in an instantaneous heat generation $E=(\sqrt{2}-1)\varepsilon s\approx 0.41\varepsilon s$ due to vortex length shortening. This heat serves as a trigger for appearance of a flux avalanche, which further development is described by the theory of thermal instability in superconductors \cite{Aranson1,Aranson2,Denisov2,Vestgarden}. As soon as $H_{e}$ reaches $H_{max}$, a usual penetration of vortices starts at flat edges as shown in Fig.\,2, the bowstring mechanism is no longer effective, and avalanches are replaced by a wide flux front penetration \cite{geom}.

It should be mentioned that Eqs.\,(\ref{eq-Hedge}) and (\ref{eq-jedge}) provide just an estimation for $H_{edge}$ and $j_{edge}$. Therefore our results described by Eq.\,(\ref{eq-full}) are not exact. However, if we introduce an unknown numerical factor $\beta \simeq 1$ in Eq.\,(\ref{eq-Hedge}): $H_{edge}\approx \beta H_{e}q\sqrt{a/s}/\pi $, then Eq.\,(\ref{eq-full}) will be modified as: $x=s\beta ^{-2/3}$, $\tan \varphi =\beta^{2/3}$ and $h=\sin (\arctan \beta ^{2/3})$. Accordingly, $H_{th}<H_{max}$ at any values of the estimation parameter $\beta $. Of course, at $\beta =1$ we come back to Eq.\,(\ref{eq-full}). Thus, our conclusion about the existence of a "bowstring window" at $H_{th}<H_{e}<H_{max}$ is valid at any reasonable estimation for $H_{edge}$. Numerical simulation for samples of various $s/a$ ratio should provide exact results for $H_{th}$.

We considered above a release of a "full" bowstring which crosses the whole edge, see line 4 in Fig.\,1. Actually penetration can start earlier when a bowstring edge tip reaches $y=s/2$ and merges with a symmetric bowstring, which propagates from an opposite corner, see lines 2 and 2* in Fig.\,1. Probability of such merging depends on the concentration of bowstrings at both corners and can vary in different experiments. If this "half-bowstring" mechanism is effective, the latter condition in Eq.\,(\ref{eq-system}) should be replaced by $x\tan\varphi =s/2$ (the rest of Eq.\,(\ref{eq-system}) remains unchanged), and penetration starts at%
\begin{equation}
\begin{split}
x_{half}=s/\sqrt[3]{4};\quad \varphi _{half}=\arctan (1/\sqrt[3]{2})\approx 38.4^{\circ };
\\
h_{half}=\sin \varphi \,\approx 0.62.   
\end{split}
\label{eq-half}
\end{equation}%
Comparing Eq.\,(\ref{eq-half}) with Eq.\,(\ref{eq-full}), we see that a "half-bowstring" effect widens the instability (avalanche) "window" $H_{th}<H_{e}<H_{max}$ but does not change qualitatively our results, see Fig.\,3. The energy (heat) effect of a half-bowstring contraction at $H_{halh}$ is $E_{half}\approx 0.61\varepsilon s$. It is worth emphasizing that $E_{half}>E$ since $\varphi_{half}<45^{\circ} $, so the total length of a half-bowstring (lines 2+2* in Fig.\,1) is greater than that of a full bowstring (line 4). Experimental results, especially obtains in strips, testify that instabilities start at $H_{th}$ rather than at $H_{half}$, see Fig.\,3.

So far we considered an isotropic superconductor. Anisotropy of "hard axis $c$" type, where $\gamma=H_{c1}^{c}/H_{c1}^{ab}=\lambda _{c}/\lambda _{ab}>1$, affects the bowstring mechanism dramatically. In most experiments with anisotropic superconductors, such as $\mathrm{MgB}_{2}$ and YBCO, epitaxial films were used, where the axis $c$ was perpendicular to the surface. The tension of an anisotropic vortex is $\varepsilon (\varphi)=\varepsilon (\sin ^{2}\varphi +\gamma ^{-2}\cos ^{2}\varphi )^{1/2}$ \cite{Balatskii,Brandt}, where $\varepsilon $ corresponds to $\varphi =90^{\circ }$. Note that $\varepsilon (\varphi )$ drops when $\varphi $ decreases, enabling easier vortex tilting and stretching across a corner. Substituting $\varepsilon (\varphi )$ into Eq.\,(\ref{eq-f}), we get the same results for $x$ and $\varphi $ as described by Eq.\,(\ref{eq-full}). It is not surprising since decreasing vortex line tension makes it easier to pull a bowstring along both axes $x$ and $y$. But for a threshold field we obtain%
\begin{equation}
h_{th}=\frac{1}{2}\sqrt{1+\frac{1}{\gamma ^{2}}}.  \label{eq-Hp-anis}
\end{equation}%
At $\gamma =1$ we get, of course, the same $h_{th}$ as determined by Eq.\,(\ref{eq-full}). We see from Eq.\,(\ref{eq-Hp-anis}) that anisotropy reduces $h_{th}$ if compare to the isotropic case: in the limit $\gamma \rightarrow \infty $ we have $h_{th}=1/2$. But the energy released due to a vortex spontaneous contraction is diminished dramatically:  $E_{anis}=(\sqrt{1+\gamma ^{-2}}-1)\varepsilon s$. For $\gamma $ substantially greater than $1$ we have $E_{anis} \approx \varepsilon s/2\gamma ^{2}\ll \varepsilon s$. An avalanche ignition is an activation process, thus such a substantial reduction of a released heat should result in complete elimination of a non-equilibrium flux penetration.

A crucial negative effect of anisotropy on a bowstring mechanism can explain an absence of spontaneous instabilities in YBCO samples with $\gamma >5$, where artificial triggering such as laser surface spotting or applying very fast external field ramping is requires.  In $\mathrm{MgB}_{2}$ avalanches are experimentally observed only at $T<10\,\mathrm{K}$. A thermomagnetic theory \cite{Aranson2,Denisov1,Denisov2} predicts the existence of a threshold temperature $T_{th}$, above which flux instabilities does not appear. However, one should take into account that penetration depth (and, correspondingly, vortex tension $\varepsilon $) in $\mathrm{MgB}_{2}$ becomes substantially anisotropic above $10\,\mathrm{K}$ \cite{Fletcher} with $\gamma \simeq 2$, which could suppress instability ignition as we discussed above.

Another experimental evidence, which supports importance of bowstrings for a non-equilibrium flux penetration, is the "crossing-field" effect \cite{Vlasko-Vlasov}.  If an in-plane (parallel to sample surface) magnetic field $H_{\parallel }$ is added to a perpendicular one in a square-shaped sample, it suppresses instabilities formation at perpendicular to the in-plane field edges, whereas an avalanche concentration at parallel to $H_{in}$ sides is no affected. This can be immediately understood within our approach, since vortices penetrating from the parallel to $H_{\parallel }$ edges just become tilted in a plane perpendicular to their direction of motion (out of plane of drawing in Fig.\,1), which has little effect. Vortices at perpendicular to $H_{in}$ are affected by an additional surface current $j_{\parallel }=\pm cH_{\parallel }/4\pi $, where the sign is different at the opposite surfaces. Adding $j_{\parallel}$ into Eq.\,(\ref{eq-jsurf}), we find that a bowstring remains inclined even after it passes a threshold position. As a result, the length contraction and, correspondingly, heat effect is diminished by a factor $\approx 1+H_{\parallel }/H_{\perp }^{c1}$, thus reducing the probability of an instability ignition.

Finally let us note that the conditions $H_{th}<H_{max}$ and/or $E \simeq \varepsilon s$ are easily broken by edge imperfections and deflection from a rectangular cross-section. Most of samples used in experiments on avalanche-like flux penetration are films, therefore it is not surprising that even in macroscopic samples with $a \simeq 1 \mathrm{mm}$ the bowstring mechanism is effective at distinct points of the sample edge. As $H_e$ exceeds $H_{max}$, edge defects become less important, and a wide flux front penetration starts.

To conclude, we described a bowstring effect for vortex penetration from the corners in flat superconductors with a rectangular  cross-section. In both disk and strip shaped samples such a penetration starts at $H_{th}$, which corresponds to $H_{edge}<H_{c1}$. Each bowstring, after it reaches the maximal length, is spontaneously contracted with an instantaneous heat release. This mechanism can be responsible for ignition (triggering) instabilities in $\mathrm{Nb}$, $\mathrm{MgB}_{2}$ and other materials. Anisotropy of penetration depth strongly diminishes the heat effect, which could explain an absence of spontaneous flux avalanches in YBCO. In general, a bowstring effect constitutes a considerable elucidation of the theory of vortex penetration into flat samples of a rectangular cross section (geometrical barrier) and is important by not only by its connection to flux avalanche triggering.

We thankfully acknowledge useful conversations with Y.\,Yeshurun and A.\,Shaulov.

\end{document}